# Performance Comparison of Dynamic Vehicle Routing Methods for Minimizing the Global Dwell Time in Upcoming Smart Cities

Tim Vranken[1], Benjamin Sliwa[2], Christian Wietfeld[2] and Michael Schreckenberg[1]
[1]Physics of Transport and Traffic, University Duisburg-Essen, Germany
e-mail: {Tim.Vranken, Michael.Schreckenberg}@uni-due.de
[2]Communication Networks Institute, TU Dortmund University, 44227 Dortmund, Germany
e-mail: {Benjamin.Sliwa, Christian.Wietfeld}@tu-dortmund.de

*Abstract*—Traffic jams in urban scenarios are often caused by bottlenecks related to the street topology and road infrastructure, e.g traffic lights and merging of lanes. Instead of addressing traffic flow optimization in a static way by extending the road capacity through constructing additional streets, upcoming smart cities will exploit the availability of modern communication technologies to dynamically change the mobility behavior of individual vehicles. The underlying overall goal is to minimize the total dwell time of the vehicles within the road network. In this paper, different bottleneck-aware methods for dynamic vehicle routing are compared in comprehensive simulations. As a realistic evaluation scenario, the inner city of Dusseldorf is modeled and the mobility behavior of the cars is represented based on real-world traffic flow data. The simulation results show, that the consideration of bottlenecks in a routing method decreased the average travel time by around 23%. Based on these results a new routing method is created which further reduces the average travel time by around 10%. The simulations further show, that the implementation of dynamic lanes in inner cities most of the time only shift traffic congestion to following bottlenecks without reducing the travel times.

## I. Introduction and Related Work

Road traffic in inner cities is often slower than on highways because the allowed speed is lower and traffic lights and intersections interrupt the traffic, which reduces the road capacity. In the past, the problem of traffic optimization was mainly addressed statically, e.g. by changing street properties, which involves cost-intensive construction work. In the context of upcoming smart cities [1] and with vehicles being equipped with wide-area communication technologies, novel methods can be applied that allow dynamic traffic flow optimization by changing the mobility behavior of individual vehicles. The topic of traffic flow optimization by trip planning is a wide field with many different approaches. One well-known attempt to describe traffic flow was done by Wardrop [2] with his two principles. In the following work "Wardrop's user Equilibrium" (following WE) will be used as the base for one of the trip planning methods. Since then multiple different approaches to a more efficient trip planning have been introduced (see [3] for a short summary and overview). A lot of the trip planning methods route every vehicle with the aim to minimize its travel costs (like travel time and travel distance for example) with the aim to achive a Nash equilibrium [4]. An example of such a routing method is presented in [5]. Because [6] showed, that a Nash equilibrium does not result in minimal average travel times, more recent works propose different approaches like [7], where routing is done dynamical and the impact of decisions is considered through a feedback of the system. This paper will take a closer look at the "Breakdown minimization Principle" [8] (following BMP), where trip planning is done based on analytic values with the goal to prevent congestion rather than reduce the travel time. The simulations will be done in a cellular automata model based off [9] and the traffic flow will be based on real-world traffic data.

To this end, first, a short explanation of how the traffic control methods are applied is done in section II followed by an introduction of an additional traffic control method as a combination of the BMP and WE. Afterwards a look at the network and the employed daily time-variation curve for the traffic flow will be taken in section III, before the trip planning methods are compared under additional constrains in section IV. In the conclusion in section V a short resume as well as an outlook on future works is given.

## II. Solution Approach

WE will be used as comparison to the other two methods because it is a very well-established method used in navigation systems. The original definition of WE is [2]:
"Traffic on a network distributes itself in such a way that the travel times on all routes used from any origin to any destination are equal, while all unused routes have equal or greater travel times."
For the static route finding this means that, each vehicle simply takes the route with the shortest travel time including the current traffic situation like jams at the moment it's created. For the dynamic route finding, this means that at beginning and every time it passes an intersection, the vehicle will be assigned the shortest route at that time step.
For the BMP, first all bottlenecks within the system have to be identified and classified. In the simulations, the system is fully disclosed and thus all bottlenecks can be obtained. The bottlenecks are classified in two different groups: traffic lights before intersections and lane merges. After identifying the bottlenecks a value $C_{\min}^k$ in $\frac{\text{vehicles}}{\text{hour}}$ has to be assigned to every bottleneck. This value is the "minimal critical value" after which traffic-breakdowns, according to the three-phase traffic theory [10], can occur, with a probability greater 0.

The size of $C_{\min}^k$ differs for lane endings and traffic light. To get an approximate value for the $C_{\min}^k$, a simple network is created. This network starts with one source track on which the vehicles are created, followed by one bottleneck (so a road with a lane merge or one that ends in a traffic light). After the bottleneck follows a track, at which end the vehicles are taken out of the system. To get a good approximation of $C_{\min}^k$ a stable flow $q^k$ of cars is sent into the system for 40 min. In those 40 min, it is monitored whether a breakdown occurs or not. The simulation is repeated 40 times to get a probability for a traffic breakdown in relation to the fixed flow $q^k$. After repeating the simulations for different values of $q^k$, a characteristic graph for this bottleneck is created.

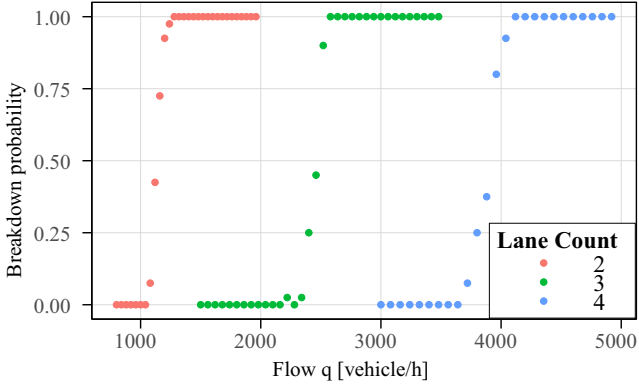

Fig. 1. Probability of a Breakdown in relation to the number of vehicle that drive over the lane ending Bottleneck.

In figure 1, the results for a simulation of a lane merge bottleneck with three different amounts of lanes before the merge is shown. It can be observed, that the probability for a traffic breakdown is always zero up to a specific $q^k$. To get an approximation for $C_{\min}^k$, either the first value of $q^k$ is chosen for which a traffic breakdown was registered or the figure is fitted according to [8]:

$$P^k(q^k) = \frac{1}{1+\exp[\beta^k(q_p^k - q^k)]}, \quad (1)$$

Where $\beta^k$ and $q_p^k$ are regression parameters. Then $C_{\min}^k$ is the value for that $P^k(C_{\min}^k) = 0$. Because $P^k(0)$ cannot be 0 with a $q^k > 0$, $C_{\min}^k$ will be chosen as the first value for that $P^k(C_{\min}^k) = 0.025 \left(\frac{1}{40}\right)$ is true. The first method is a lot faster since the simulation has to be only executed until the first breakdown is registered and then $C_{\min}^k$ is set as the $q^k$ for which it was registered, but because the second way uses more information to reduce the error it is applied to get the values for $C_{\min}^k$. For lane merges, $C_{\min}^k$ is given by:

$$C_{\min}^k \approx 1000 + 1350 \cdot (N_s - 2) \, \frac{\text{Vehicles}}{\text{h}}, \quad (2)$$

Where $N_s$ is the number of lanes before the lane ends. Note that that the maximum speed seems to have no impact on $C_{\min}^k$ for the used model of lane merging bottlenecks since the vehicles reduce their speed significantly on both lanes to allow the merging.

For the traffic light bottlenecks, $C_{\min}^k$ is given by (in consistence with [11], [12]):

$$C_{\min}^k \approx q_{\text{sat}} \cdot \frac{G^{\text{eff}}}{Ap} \quad (3)$$

Here $Ap$ is the time of one period of the traffic light. $q_{\text{sat}}$ is the number of vehicles that can pass over one green phase (in $\frac{\text{Vehicles}}{\text{h}}$) and $G^{\text{eff}} = Ap - Y - R - \Delta t$ is the effective green time of the traffic light. $R$ is the red time and $Y$ the yellow time of the traffic light. According to [12], the effective green time of a traffic light is given by $G^{\text{eff}} = G \cdot \Delta t$, where G is the real green time of the traffic light and $\Delta t$ is $\approx 3-4$ s. The values for the simulations are chosen in such a way that the average time a vehicle needs to drive over a green light after the driver in front passed the light needs $\approx 2.5$ s in consistence with [13]. This results in values of $\Delta t = 3.5$ s and $q_{\text{sat}} = 1682 \, \frac{\text{Vehicles}}{\text{h}}$ for straight leading lanes in front of traffic lights. The average traffic flow is reduced to $q_{\text{sat}} = 1469 \, \frac{\text{Vehicles}}{\text{h}}$ for right and left turn lanes before traffic lights because the average speed before a turn is lower than in free flow. Note that this $C_{\min}^k$ for traffic lights is peer lane and peer direction before the traffic light. Therefore, a traffic light with one right turning lane, three straight leading lanes and two left turning lanes would count as three bottlenecks with the respective values. Additionally, lanes that allow drivers to drive straight ahead and turn right or left, also have to be differentiated in separated bottlenecks from lanes where only one of these degrees of freedom is present. This way, traffic flow that will turn right/left on this lane can be weighted greater than the straight driving flow to take the slower speed of vehicle that turn into account.

Lastly, for right turning lanes that are not restricted by a traffic light (non-signalized turn lanes) additionally an approximation for the period in that vehicles on this lane can cross the intersection have to be determined. Since $C_{\min}^k$ is the minimal value for which a breakdown can happen, it can be assumed, that on the roads that lead to an interaction with this right turning lane, the amount of vehicles is equal to the amount that can pass one green light phase. With this assumption, a theoretically green time for the lane can be found as:

$$G^{\text{theo}} = Ap - \sum^{i} G_i^{\text{eff}} - \Delta t \; . \quad (4)$$

Here, the sum is over all traffic lights $i$ at the intersection that lead to the same lane as the right turning lane. For obtaining $C_{\min}^k$ for those kind of lanes $G^{\text{eff}}$ is replaced with $G^{\text{theo}}$. Tests showed, that further $\Delta t$ has to be changed to 4.8 since after the other lanes green phase ended the vehicles still have to clear the road before the first car can start.

After setting all the values, the traffic flow will then be directed in such a way that the flow $q^k$ at no bottleneck $k$ is equal or greater $C_{\min}^k$. If the traffic flow is so high that this condition cannot be guaranteed anymore, the flow $q^k$ should not be greater than $C_{\min}^k + \epsilon^k$ where $\epsilon^k$ is the smallest value possible. A simple example for this would be a road that is divided in two different directions. After each of them passes a bottleneck, the roads are reunited again. According to the

BMP the equation system:

$$q^1 + s = C_{\min}^1 \quad (5)$$
$$q^2 + s = C_{\min}^2 \quad (6)$$
$$q^1 + q^2 = q_{\text{sum}} \quad (7)$$

has to be solved with $q_{\text{sum}}$ being the total flow of cars starting at this road (in $\frac{\text{Vehicles}}{\text{h}}$). To ensure that the probability of a breakdown is minimal, the solution with the highest value of s is chosen. When the Value of $s$ reaches 0, this means that the traffic flow reached the critical value at at least one bottleneck. Then $\epsilon^k$ has to be added at every bottleneck where $q^k >= C_{\min}^k$. From then on, $s$ will is not maximized at the bottlenecks but $\sum^k \epsilon^k$ is minimized. Afterwards, every vehicle gets the route $r^k$ assigned that leads over the bottleneck $k$ with the probability $\frac{q^k}{q_{\text{sum}}}$. Through these equations it can be seen, that the travel times of the routes are not considered. This could result in vehicles selecting the longest route possible even at very low traffic volumes. To prevent vehicles from taking routes with to long travel times (at free flow) a preselection of the routes that are used by the BMP has to be made. For this preselection the travel times of all routes at free flow are determinated and then only routes with a trip duration up to $\tau + \delta$ — where $\tau$ is the travel time of the shortest route and $\delta$ is a system specific time that has to be defined— are considered.

Lastly, an optimization method based on WE and the BMP, which brings together the beneficial aspects of both of them, is created. This method works in two steps. First, for every vehicle at the time step of its creation the shortest route (based on the current travel time) is determined. For this route it is checked whether

$$q_{\text{as}}^k(t_{\text{ex}}^k) + 1 + s(t_{\text{ex}}^k) < \frac{C_{\min}^k \cdot \Delta t_{\text{ex}}^k}{60} + \epsilon^k(t_{\text{ex}}^k) \quad (8)$$

is fulfilled for every bottleneck on the route or not. Here $q_{\text{as}}^k(t_{\text{ex}}^k)$ is a counter that is increased by 1 for every bottleneck on the route of the vehicle as soon as it is created. $t_{\text{ex}^k}$ is the expected time of arrival at the bottleneck $k$ rounded to a $\Delta t_{\text{ex}}$-minute time frame. $s(t_{\text{ex}}^k)$ and $\epsilon^k(t_{\text{ex}}^k)$ have the same function as in the BMP, only restricted to the time frame $\Delta t_{\text{ex}}$. To adapt $C_{\min}^k$ to the $\Delta t_{\text{ex}}$ minute interval it is multiplied with it and then divided by 60 (to get from $\frac{\text{cars}}{\text{hour}}$ to $\frac{\text{cars}}{\text{minute}}$). If equation (8) is not fulfilled for one or more of the bottlenecks on the route, the next shortest route is tested and so on until the shortest route on which the condition is fulfilled for every bottleneck is found. If none of the possible routes fulfills the condition, $s(t_{\text{ex}}^k)$ is decremented by one or, when $s = 0$ than, $\epsilon^k$ is increased by one at the bottleneck with the greatest $\epsilon^k(t_{\text{ex}}^k)$. Afterwards the equations are checked again beginning from the one with the shortest route. For the simulations, $\Delta t_{\text{ex}^k} = 6 \cdot \frac{85}{60}$ min as well as the starting value of $s(t_{\text{ex}}^k) = 1$ showed the best results (the traffic signal cycles time frame for all traffic lights in the simulations is 85 s). $s(t_{\text{ex}}^k) = 1$ showing the best results is an indicator that the $C_{\min}^k$ values are a good approximations because the traffic flow can get close up to $C_{\min}^k$ without creating breakdowns which would slow down the cars more than it would take them to select the next shortest route. Because vehicles are still send over long routes to prevent congestion at high traffic volumes, a preselection of routes still has to be made to prevent the use of routes that take too long.

## III. NETWORK AND DAY TIME-VARIATION CURVE

Before applying the different optimization methods, the network used for the comparison is introduced that is based on the inner city main roads of Duesseldorf. The used roads together with the intersections are shown in Figure III in blue.

The system has $a \in \mathbb{N} \land 0 < a < 11$ streets which serve as a source (marked green) and $b \in \mathbb{N} \land 0 < b < 11$ streets that serve as sinks (marked red).

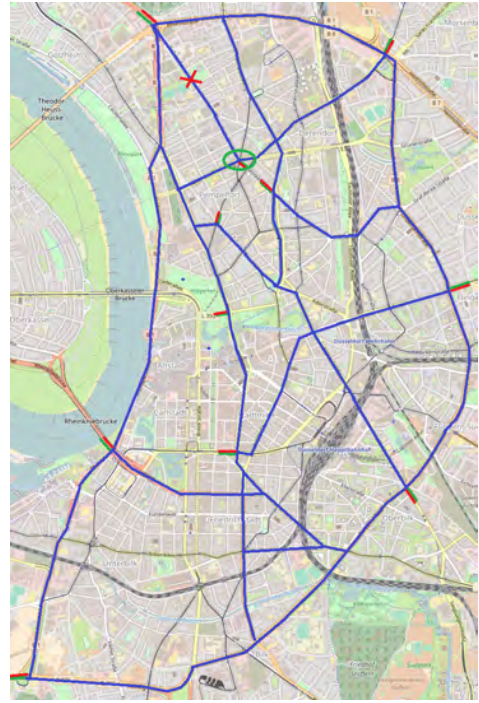

Fig. 2. Marked area in the Duesseldorf inner city. The blue marked streets and intersections are streets used in the simulation. The green marked streets act as sources and the red marked streets are sinks. The streets marked with a red x have a road closed in a later simulation. The intersections marked with a green circle have dynamic lanes that change in an additional simulation based on the time of the day to adjust to traffic demand. (Map: ©OpenStreetMap contributors, CC BY-SA)

For the creation of realistic traffic flow behavior, the data from a video detector placed on the "Voelklinger Straße" in Duesseldorf in the northern direction is used. This detector periodically reports the number of cars that have passed it as well as their velocity each 60 s. These values are stored, extrapolated to $\frac{\text{Vehicles}}{\text{h}}$ and then displayed in figure 3 a) for the dates from 01.03.14 till the 30.04.14 (for two months) with the blue line marking the average traffic flow registered. Note, that the detector counts the vehicles on two lanes only as a single value, meaning the average traffic flow over one lane is half the value in 3 a). Based on this blue line, the day time-variation curve shown in figure 3 b) is created. The traffic flow between the morning and the midday high-phase is purposely made even lower to get a more characteristic line where the two peaks are easily distinguishable. For the same reason, the midday high-phase value is increased. This function now serves as a factor that can be multiplied by the maximum number of vehicles that drive over the source lane, so $C_{\min}^k$ if the source lane ends in a bottleneck and $1700 \frac{\text{Vehicles}}{\text{h}}$ per lane (maximum spawn rate within the simulation) else. Additionally, two more factors have to be applied. One factor takes into consideration how many lanes the sink street consists of, since sinks have more lanes if they are more frequently used. This is done by dividing the number of lanes $S_b$ of the sink $b$ by the number of lanes of all sinks combined ($S$). Lastly, a factor is added to weight different directions for the traffic. Duesseldorf is a big commuter city [14]. This means that in the morning more commuters drive into Duesseldorf to work than leave the city to work in nearby areas. After work at the midday high-phase, the behavior is reversed. Now more commuters leave Duesseldorf than drive back into it. The number of incoming commuters is also greater than the number of people living and working there. Lastly, the number of drivers that pass through Duesseldorf (so are starting on the outside ring and drive to another sink on the outer ring) are the fewest.

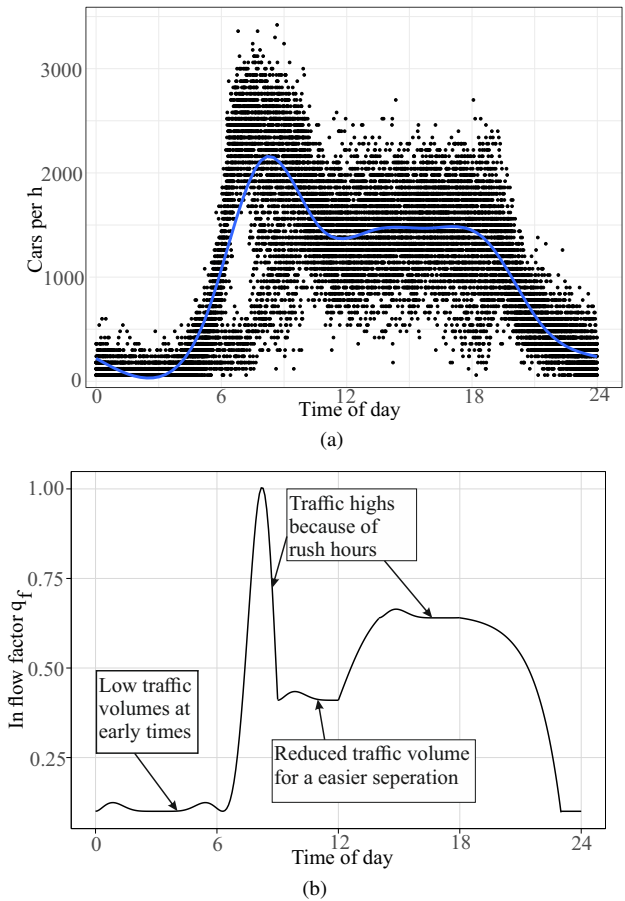

Fig. 3. a) Average cars per hour at different times of the day for a street with two lanes in Duesseldorf over two month obtained from real-world detectors. The Blue line marks the average traffic flow at the time. b) Time-variation curve that is artificially created with two clearly separable highs. Fig. a) was created as part of the bachelor thesis "Empirische Untersuchung von Verkehrsmustern in übersättigtem Verkehr." of Cederic Wilting at the University Duisburg-Essen.

For accounting these differences, the additional factor $(g(t)_{ab})$ is introduced. Combining all the factors, the resulting traffic flow from source $a$ to sink $b$ is defined by:

$$Q_{ab} = \sum_{k=i}^{j} \min(C_{\min}^k, 1700) \cdot g(t)_{ab} \cdot S_a^k \cdot S_b/S \ \frac{\text{vehicle}}{\text{h}}. \quad (9)$$

The sum over $k$ is $a$ for all bottlenecks at the end of the source street. For example, a source lane that ends in a traffic light with one right-turning lane and two straight leading lanes would be the sum over those two bottlenecks. For simplicity, the incoming traffic volume is kept constant (with stochastic stagger) for 10 simulated minutes after which it is updated. With this method, traffic in the network can now be created for comparing the considered optimization methods. 3 different values are compared for each method:

1. travel times through the system.
2. number of Breakdowns in the system.
3. length of breakdowns in the system.

For obtaining the average travel time, floating car data [15] are used. This means, that the travel time of every car is recorded in the time step when the car reaches its destination and is taken out of the system. In order to obtain the number of breakdowns and the length of the breakdowns, induction loops were created 300 m before every bottleneck. These induction loops register the number of cars and their respectivy velocity. Once a minute, the number of cars and their average speed is reported. A breakdown is registered, in accordance with the rules from [16], if the average speed drops below $30 \frac{\text{km}}{\text{h}}$ and stays below it for at least 15 min. The breakdown then ends when the average speed rises above $30 \frac{\text{km}}{\text{h}}$ again and stays there for over 20 minutes.

Before the comparison of the methods is provided, first it is tested how well the theoretical $C_{\min}^k$ values of chapter II fit the simulation. Therefore, the $C_{\min}^k$ are increased by $100 \frac{\text{Vehicles}}{\text{h}}$ and then traffic flow evaluations applying the BMP are performed for 100 days. During the simulations, the $C_{\min}^k$ value is decreased every time a breakdown happens with $q^k < C_{\min}^k$. Figure III shows the differences between the theoretical and the simulation $C_{\min}^k$ for every day. It can be seen that the average difference between theoretical value and simulation decreases with time. This means that the theoretical value is at least not higher than the effective ones. The average distance not decreasing below $88 \frac{\text{vehicle}}{\text{h}}$ would imply the theoretical value being too low, but actually the value of only a few bottlenecks (13 of the 160) are decreased over time. For the other bottlenecks, the traffic flow does not enter the interval $C_{\min,\text{theo}}^k < q^k < C_{\min,\text{sim}}^k$ between the theoretical and the increased simulation value. This implies that the BMP is stable under small fluctuations in the traffic volume and the applied values for $C_{\min}^k$. The 13 bottleneck, for which the traffic flow is within this interval, decrease by $100 - 120 \frac{\text{Vehicles}}{\text{h}}$. This means that the value $C_{\min}^k$ fit the simulations quite well (the overestimation of $20 \frac{\text{Vehicles}}{\text{h}}$ is within the stochastic staggering).

## IV. RESULTS

After confirming the value of $C_{\min}^k$, the simulations can be performed and evaluated. For visualization, the travel times (as an average over 10 min) as well as the $95\%$ confidence interval are shown in figure IV. In the following figure 6, the results for the three observables as mentioned above are shown in histograms of their distributions for 100 simulated days. Afterwards, the simulations are repeated two times, where each time a change was performed within the network. For the red values in figure 7, dynamic lanes where implemented which are used to change the directions based on the time of the day to support the changing demand at different times of the day due to commuters. In the simulations of the blue values for figure 7, a lane of the street marked in figure III is temporarily closed from 8.40 o'clock till the end of the day to simulate roadwork.

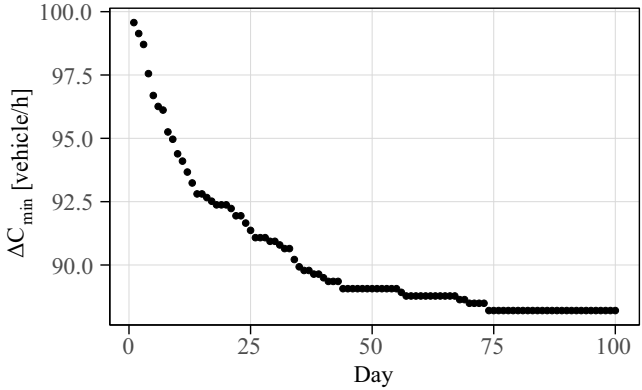

Fig. 4. Average difference of the theoretical $C_{\min}^k$ and the one used for the simulation within the simulation over 100 Days.

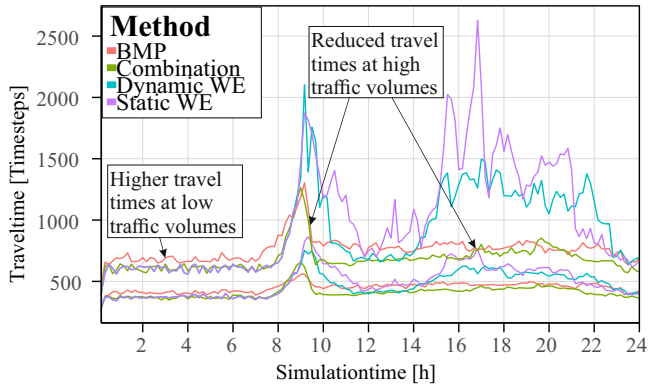

Fig. 5. Travel times for the four different traffic assignment methods. The lower lines represent the average travel time taken over 10 min while the upper line represents the 95% confident interval upper bound.

In figure IV, different key characteristics can be observed. During the early times of the day, when the traffic volume is low, the BMP results in higher travel times than the other methods. This is caused by the BMP not trying to minimize travel times but the number of breakdowns. The distribution of the traffic flow (like in the example with equation (5) to (7)) is done without considering the travel times of the route $q^n$, resulting in vehicles taking longer routes than necessary. At times of high traffic volumes, the BMP optimizes better than the static and the dynamical WE because all cars, that are routed based on WE, drive over the same route, which will always result in jams as soon as their amount is exceeds the road capacity. The next vehicle that enters system and is routed will then take the next shortest route, which again results in a jam on this route and so on until the first jam is dissolved. The dynamic WE only chances the frequency at which the routes are changed, but because vehicles which would have arrived at the jam after it was dissolved are also rerouted, the advantage of the dynamic rerouting actually dissolves resulting in no decrease for the average travel time. The combination has lower travel times than the BMP early in the morning and in the afternoon. Only the morning peak is high enough to cause jams, which increase the average travel time above the one of the BMP shortly.

Figure 6 a) shows the same results as IV but for 100 simulated days and as a distribution of travel times. Again, one can see that (static and dynamic) WE creates higher travel times while the combination of WE and the BMP results in even smaller travel times than the BMP. Additionally, in 6 b) the distribution of jam lifetimes is shown. Here it is clear, that the BMP creates far less breakdowns than the other methods. Even the combination of the BMP and WE creates more breakdowns than the BMP since not only the breakdown probability is minimized but also the travel time. This means that the bottlenecks on the shortest routes are used up to $C_{\min}^k$ while the other bottlenecks are still unused compared to the BMP where the traffic flow is distributed evenly over all bottlenecks. Compared to WE the number of breakdowns is still significant lower and as one can see in the travel times, the increased number of breakdowns is less important for the average than the weighting of routes based on their travel times.

The red values in figure 7a show the changes to the travel times and jam lifetimes when applying dynamic lanes to the intersection. For the BMP, a reduction in the number of breakdowns for nearly all jam lifetimes can be observed. This reduction results in a shift of the travel times to shorter values. Since the number of breakdowns in the system was already low compared to the other trip planning methods, the impact of the dynamical lanes on the global dwell time is low (a reduction of 0.2% in the travel time values as seen in table I). Compared to the BMP, the changes in the dynamic WE are significant greater since the remaining margin left for improvements is bigger. As it can be observed, the value of the travel time for the dynamic WE in table I is reduced by around 4.9%. The results of the static WE show that here the dynamical lanes do not result in a reduction of breakdowns but shift the jam lifetimes to shorter values. This means that an increased traffic volume, which still causes the breakdown, compensates the additional lane at that time. As soon as the traffic before the bottlenecks breaks down, the new incoming traffic volume is significantly reduced and the additional lane helps to resolve the jam. When looking at the travel time, changes in the traffic in the morning and the afternoon can be observed. The shift of the travel times below 500 s to the values one-step lower is the change that is coming from the morning traffic while the bigger shift of the travel times greater 500 s to smaller values comes from the traffic in the afternoon. The impact of the dynamic lanes in the morning is smaller (only reduces the travel time by around 3.6% compared to the 11.2% in the afternoon), because here the additional traffic volume that passes the bottleneck cannot be supported by the traffic capacity of the following traffic. Lastly, the combination of the BMP and WE is not able to make use of the dynamic lanes. The breakdowns are not reduced and only shifted to smaller values comparable to WE while the impact on the travel times is smaller than the statistic staggering.

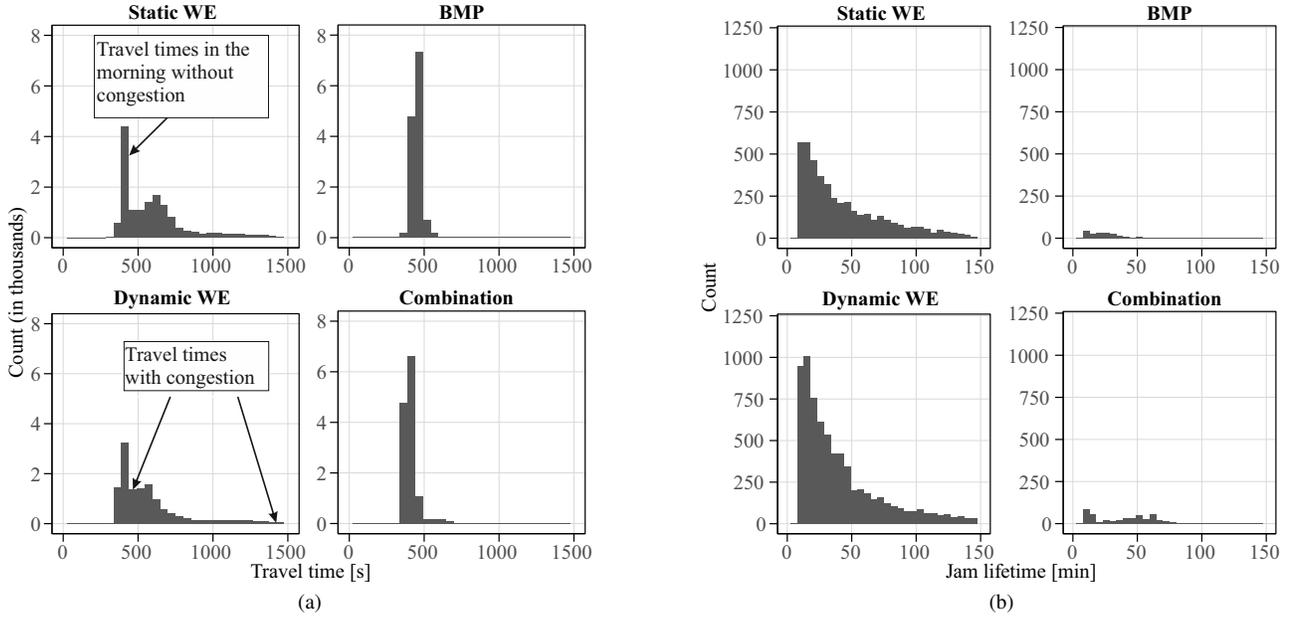

Fig. 6. Simulation results for 100 simulated days. a) Distribution of the over 1 min averaged travel times for the different routing methods. b) Distribution of the jam lifetimes for the different routing methods.

To conclude the simulations, 100 days were simulated, where a lane was closed at 8.40 o'clock. The changes of introduced by this simulation are shown in figure 7 in blue. It can be seen that the differences for the BMP as well as for the combination are even smaller than for the dynamical lanes and non-significant. Surprisingly, a significant improvement for the travel times while applying the static WE is found. This behavior can be explained by considering the following intersection which is the same one at which the dynamical lanes were tested. There, the additional lane provided a significant improvement for the travel time. This means that big jams were created before this intersection. The closed lane now lies before the intersection, so that the jam is created earlier and more vehicles take alternative routes with shorter travel times, which improves the average travel time. The earlier creation of — and thus reaction of the driver to — the jams explains the shift in the lifetimes of jams to shorter ones. The dynamical WE on the other hand, overcompensates the jam. Here, the vehicles avoid the jam so early that they choose routes where jams would have been created even without the additional traffic flow. The jams on the alternative routes involve more vehicles and thus take longer to dissolve, which results in increased average travel times.

## V. CONCLUSION

In this paper, traffic flow of an inner city network was simulated with different trip planning methods to analyze their effect on the overall dwell time as well as congestion lifetime in different situations. Traffic volumes in inner cities often exceed the road capacity while the road network is limited in space most of the time. Consequently, increasing the traffic capacity through building additional lanes is rarely possible or not desired because of the involved costs. To satisfy the increasing demand in traffic volume, dynamical lanes that change the traffic capacity at bottlenecks depending on the time of day were evaluated, different trip planning methods were analyzed on their efficiency and a new trip planning method was introduced. The simulations were able to show that the new trip planning method resulted in a significant reduction of the dwell time at high and low traffic volumes while the other methods only optimized the traffic behavior for either high or low traffic volumes. Simulations with dynamic lanes at only one intersection in the city showed that in most cases, only the locations of the breakdowns are shifted to the next bottlenecks, without a positive system-wide effect on the average dwell time. Only the increased capacity in the afternoon leading out of the city, could be used efficiently since traffic capacity outside of cities is often comparable higher. In future work, the additional time vehicles need to pass a congestion, based on the additional flow $\Delta t_{ex}^k$ over the respective bottleneck will be approximated. With this additional value, vehicle can be routed not only based on the current travel time but also with considering how the traffic flow will behave in the future. This will shift the goal of minimization more from the breakdowns to the travel times since vehicles can be routed through bottlenecks even if it increases the risk of a breakdown, as long as it has a positive impact on the dwell time. With this change, the combination method no longer needs a preselection of routes since routes that take longer than the travel through congestion will not be chosen. Furthermore, other dynamic options like dynamic traffic lights, with which not only the traffic capacity at one bottleneck can be changed, but also at the following bottlenecks, will be analyzed. These two options will both be done considering the goal of a reduction of the global dwell time as well as an increased accuracy of the description of the traffic flow.

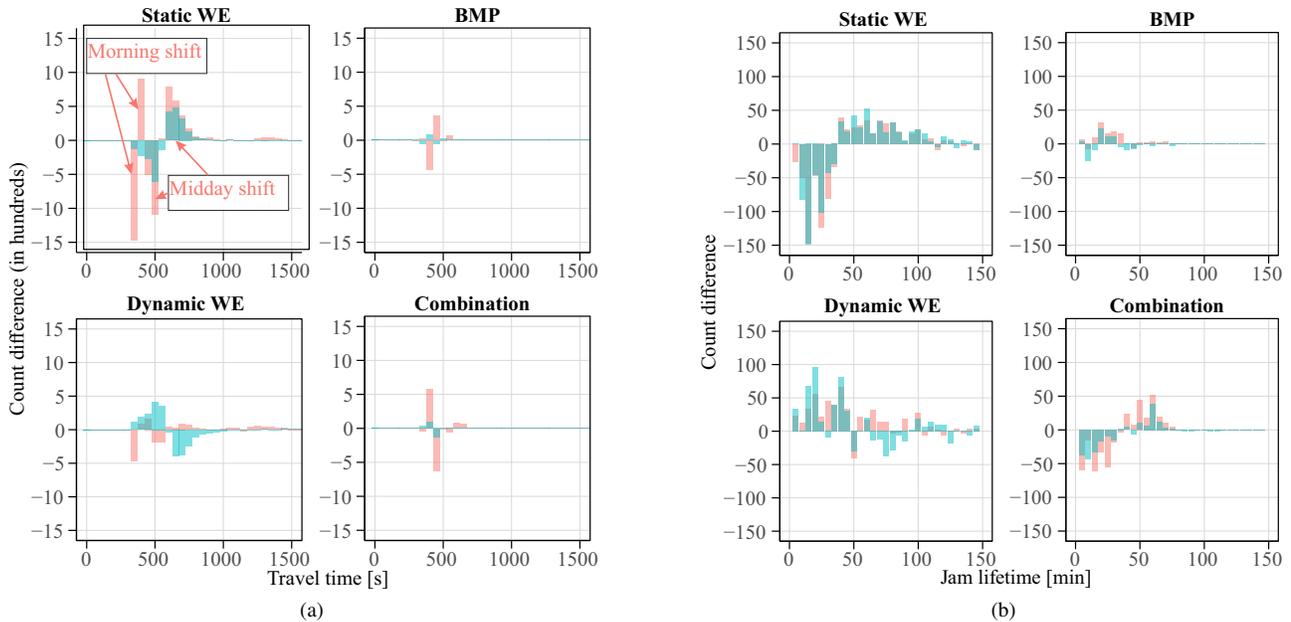

Fig. 7. Differences between the normal simulation and the simulations with changes. Positive count differences mean that these values appeared more often in the normal simulation while negative ones show how much more often these values appeared in the simulations with changes. The red values represent the difference to the simulations with dynamic lanes while the blue ones mark the difference to the simulations with the closed lane. a) Count difference of the over 1 min averaged travel times for the different routing methods in hundreds. b) Count difference of the jam lifetimes for the different routing methods.

TABLE I
AVERAGE TRAVEL TIME $\bar{\tau}$ AND THE STANDARD ERROR $se$ FOR ALL FOUR METHODS AND THE THREE SIMULATIONS.

| Method | $\bar{\tau}$ normal | $se$ normal | $\bar{\tau}$ Dynamical lanes | $se$ Dynamic lanes | $\bar{\tau}$ lane closing | $se$ lane closing |
|---|---|---|---|---|---|---|
| Static WE | 589.2 | 1.9 | 545.1 | 1.8 | 568.96 | 2 |
| Dynamical WE | 590 | .23 | 561.3 | 2 | 610 | 2.2 |
| BMP | 449.7 | 0.3 | 448.2 | 0.3 | 448.9 | 0.3 |
| Combination | 408.9 | 0.4 | 409.2 | 0.4 | 410 | 0.5 |

ACKNOWLEDGMENT

Part of the work on this paper has been supported by Deutsche Forschungsgemeinschaft (DFG) within the Collaborative Research Center SFB 876 "Providing Information by Resource-Constrained Analysis", project B4.